\documentclass[lettersize,journal]{IEEEtran}

\usepackage{configuration}

\title{Enabling Real-Time Programmability for RAN Functions: A Wasm-Based Approach for Isolated and Deterministic dApps}

\author{

    \IEEEauthorblockN{João Paulo Esper, Yure Freitas, Pedro Souza, Bruno Silvestre,\\ Joao F. Santos, \textit{Senior Member, IEEE}, Alexandre Huff, Cristiano Both, and Kleber Cardoso}

    \thanks{
    
    João Paulo Esper, Yure Freitas, Bruno Silvestre and Kleber V. Cardoso are with the Universidade Federal de Goiás (UFG), Brazil; Pedro M. Souza and Cristiano B. Both are with the Universidade do Vale do Rio dos Sinos (UNISINOS), Brazil; Joao F. Santos is with the Commonwealth Cyber Initiative (CCI) and Virginia Tech (VT), USA; Alexandre Huff is with the Universidade Tecnológica Federal do Paraná (UTFPR), Brazil.}

}

\begin{document}

\bstctlcite{IEEEexample:BSTcontrol}

\maketitle

\begin{abstract}
The \ac{O-RAN} architecture enables third-party applications to optimize \acp{RAN} at multiple timescales. However, the real-time distributed applications (known as dApps) that demand low latency, high performance, and resource isolation remain underexplored. Existing approaches propose collocating a new \ac{RIC} at the edge, or deploying dApps in bare metal along with \ac{RAN} functions. The former approach increases network complexity and requires additional edge computing resources, and the latter raises serious security concerns due to the lack of native mechanisms to isolate dApps and \ac{RAN} functions. Meanwhile, \ac{Wasm} has emerged as a fast, lightweight technology for robust execution of external, untrusted code. In this work, we propose a new approach to executing dApps using \ac{Wasm} to isolate applications in real-time in \ac{O-RAN}. Results from a proof-of-concept implementation show that our lightweight and robust approach can provide predictable, deterministic performance, resource isolation, and low latency, supporting real-time control loops.
\end{abstract}
    

\acresetall

\iftrue
\fancypagestyle{firstpage}
{
    \fancyhead[L]{This work has been submitted to the IEEE for possible
      publication.\\
      Copyright may be transferred without notice, after which this version may no longer be accessible.}
    \fancyhead[R]{}
}
\thispagestyle{firstpage}
\fi

\section{Introduction}\label{secI}

The architecture of the \ac{O-RAN} decomposes the \acp{BS} into discrete functional components: the \acp{CU}, \acp{DU}, and \acp{RU}, each implementing different layers of the 5G protocol stack. This disaggregation enables \acp{MNO} to deploy, scale, and control individual \ac{RAN} functions based on user demand, network topology, and operational costs. In addition, \ac{O-RAN} introduces two orchestrators for managing \ac{RAN} functions, known as the \acp{RIC}, which can run third-party applications for monitoring, controlling, and optimizing several aspects of the \ac{RAN} on different timescales~\cite{polese2023understanding}. The \ac{Non-RT RIC} hosts rApps, implementing long-term tasks on timescales above \unit[1]{s}, e.g., data analytics, \ac{AI}, and \ac{ML} model training~\cite{santos2025managing}. In contrast, the \ac{Near-RT RIC} hosts xApps that implement time-sensitive tasks operating at shorter timescales between \unit[10]{ms}-\unit[1]{s}, e.g., radio resource management, load balancing, and traffic steering. 

\begin{figure}[t!]
    \centering
    \includegraphics[width=1\linewidth]{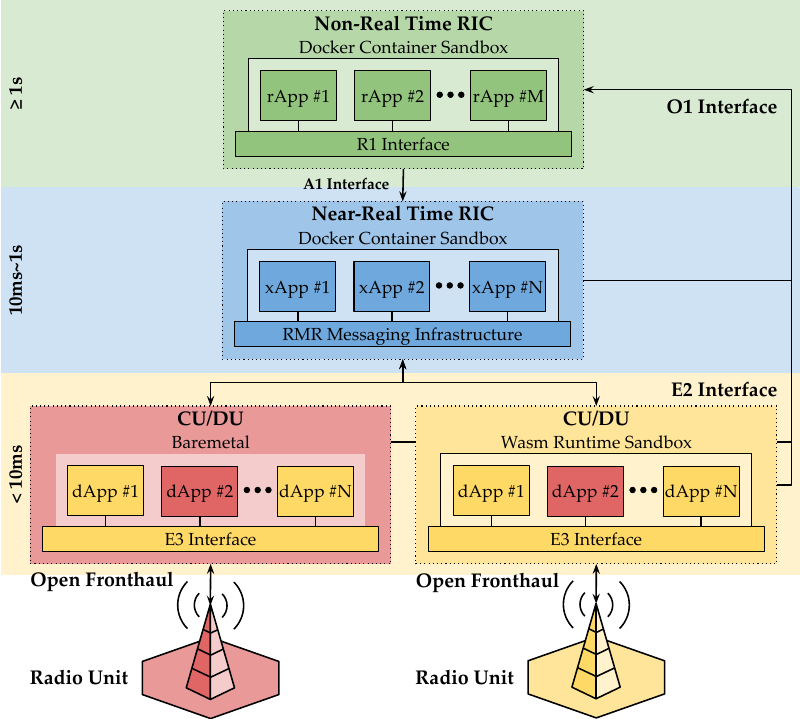}
    \caption{
    Without mechanisms to ensure robust operation of third-party applications in \ac{O-RAN}, failures or misbehavior in a single application can compromise not only an individual RAN function but the entire chain of dependent functions.}
    \label{fig:o-ran-wasm-dapp-arch}
\end{figure}

There are several real-time decisions and optimizations that occur on even lower timescales, below \unit[10]{ms}, e.g., scheduling, beam management, and spectrum sensing, which are considered for further studies in the \ac{O-RAN} specifications, but lack support in current standards~\cite{d2024dapps, d2026dapps}. This constraint has led to several initiatives to extend the \ac{O-RAN} architecture to support real-time control loops~\cite{dapps2022,listenwhiletalking,lacava2025}, introducing a new type of third-party applications for implementing low-latency tasks and real-time control loops, known as \textit{dApps}.

At such low timescales, a large portion of the delay in control loops is attributed to the communication latency between the \ac{BS} and \acp{RIC}, which can reach dozens of milliseconds~\cite{bruno2024evaluating}. Therefore, the location of the \acp{RIC} becomes a limiting factor for real-time control loops and programmability of the \ac{RAN}. To address this issue, some works have proposed introducing a \ac{RIC} collocated with \acp{CU}/\acp{DU} in the edge, as a new \ac{RT-RIC} designed to run real-time applications~\cite{edgeric,edgericdemo,cliutinyric,tinyricml}. While this approach reduces communication latency and enables real-time programmability, it increases network complexity and assumes the availability of computing resources to host an entirely new \ac{RIC}, which may not always hold, as edge computing resources are usually scarce and expensive~\cite{bruno2024evaluating}. Other approaches propose deploying dApps that interact directly with \ac{RAN} functions~\cite{lacava2025,dapps2022,listenwhiletalking}. However, running third-party applications on bare metal raises serious security concerns, as there are no native mechanisms to guarantee isolation between dApps, or mitigate the detrimental impact of misconfigured, faulty, or malicious dApps controlling \ac{RAN} functions, as illustrated with \textit{dApp \#2} in Fig.~\ref{fig:o-ran-wasm-dapp-arch}.

Ensuring the robust operation of third-party applications in the presence of faults and failures is a known challenge in \ac{O-RAN}, which currently relies on containerization to isolate xApps and rApps running in the Near- and \acp{Non-RT RIC}, respectively~\cite{santos2025managing}. However, traditional containers introduce virtualization overhead for executing applications in a sandbox environment (detailed further in Section~\ref{sec:back}). Meanwhile, \ac{Wasm} has recently emerged as a lightweight technology for fast, robust execution of external, untrusted code. Originally designed for Web browsers, \ac{Wasm} is gaining traction as a general-purpose sandbox that supports the deployment of third-party applications in multiple languages with near-native performance, while providing strong fault detection and isolation guarantees~\cite{menetrey2022webassembly}.

In this paper, we propose a new approach to secure the operation of dApps in \ac{O-RAN} by using \ac{Wasm} to isolate third-party applications in real-time. Results from our proof-of-concept implementation show that our lightweight and robust approach can provide predictable, deterministic performance for real-time control loops. Our contributions are as follows:

\begin{itemize}

\item We detail dApp standardization efforts, assess existing approaches for creating dApps, and compare them based on architectural complexity and security capabilities.

\item We propose \ac{Wasm} as a new lightweight sandbox for isolating dApps from one another in \ac{O-RAN}, providing isolation and deterministic performance, while both running on top of existing dApps frameworks and avoiding the introduction of additional \ac{RIC} layers.

\item We created a reference \ac{Wasm}-based dApp to validate \ac{Wasm}'s fine-grained isolation and compared its performance against containers and bare-metal deployments in terms of control-loop latency and computational footprint.

\end{itemize}

Section~\ref{sec:back} discusses related work. Section~\ref{sec:wasm} introduces \ac{Wasm} and its opportunities in \ac{O-RAN}. Section~\ref{sec:poc} details our prototype, Section~\ref{sec:exp} presents experimental results, and Section~\ref{sec:conc} discusses open challenges.
\section{Background and related work}\label{sec:back}

The current literature proposes different approaches for creating third-party applications and enabling real-time programmability of the \ac{RAN}, which we refer to as dApps, following the nomenclature adopted in the recent O-RAN technical report~\cite{d2024dapps}. The key distinction among these approaches lies in the environment in which dApps execute and in their robustness mechanisms, as illustrated in Table~\ref{tab:comparisons}.

\begin{table}[!t]
\centering
\resizebox{\columnwidth}{!}{%
\begin{tabular}{c|c|c|c|c}
\rowcolor[HTML]{CFE2F3} 
\cellcolor[HTML]{CFE2F3}\textbf{dApp} & \multicolumn{2}{c|}{\cellcolor[HTML]{CFE2F3}\textbf{Deployment Locations}} & \multicolumn{2}{c}{\cellcolor[HTML]{CFE2F3}\textbf{Robustness Mechanisms}} \\
\rowcolor[HTML]{CFE2F3} 
\cellcolor[HTML]{CFE2F3}\textbf{Proposal} & A New \ac{RT-RIC} & CU/DU & Isolation & Metering \\ \hline \hline
\rowcolor[HTML]{9FC5E8} 
\cite{edgeric} & \checkmark &  & \checkmark & \\
\rowcolor[HTML]{CFE2F3} 
\cite{edgericdemo} & \checkmark &  & \checkmark & \\
\rowcolor[HTML]{9FC5E8} 
\cite{cliutinyric} & \checkmark &  & \checkmark &  \\
\rowcolor[HTML]{CFE2F3} 
\cite{tinyricml} & \checkmark &   & \checkmark &  \\
\rowcolor[HTML]{9FC5E8} 
\cite{dapps2022} &  & \checkmark &   &  \\
\rowcolor[HTML]{CFE2F3} 
\cite{listenwhiletalking} &  & \checkmark &   &  \\
\rowcolor[HTML]{9FC5E8} 
\cite{lacava2025} &  & \checkmark &   &  \\
\rowcolor[HTML]{CFE2F3} 
\textbf{This Work} &  & \checkmark &  \checkmark  & \checkmark
\end{tabular}%
}
\caption{Comparison\,between\,different\,approaches\,to support dApps 
and\,real-time\,programmability\,for\,\ac{RAN} functions regarding
their deployment\,locations\,and\,robustness\,mechanisms.}
\label{tab:comparisons}
\vspace{-2em}
\end{table}

EdgeRIC~\cite{edgeric,edgericdemo} and TinyRIC~\cite{tinyricml,cliutinyric} introduce new orchestrators deployed at the edge, collocated with or near the \ac{RAN} functions. This approach resembles the creation of the Near- and \acp{Non-RT RIC} in \ac{O-RAN}, which hosts xApps and rApps. However, unlike these \acp{RIC}, which are typically deployed in regional or central clouds with ample computational resources, the edge is resource-constrained, with limited and costly compute capacity. The works of \cite{lacava2025, listenwhiletalking, dapps2022} propose a more streamlined approach, where dApps interact directly with \acp{CU}/\acp{DU} through a single new interface that exposes \ac{RAN} metrics and controllable parameters in a highly efficient, low-latency manner. While this design simplifies deployment and reduces overhead, it lacks a native sandbox environment for executing dApps.

Near- and \acp{Non-RT RIC} employ containerization to create sandbox environments for isolating third-party applications, with well-defined APIs to leverage the capabilities of the \acp{RIC} for controlling the \ac{RAN}, along with tools for logging, collecting metrics, and triggering alarms in case of failures~\cite{santos2025managing}. While this approach enables robust operation of third-party applications, it also introduces virtualization overhead due to the abstraction of system libraries, package dependencies, and the network stack to execute applications in a sandbox environment~\cite{wiegratz2024comparing}. Meanwhile, \ac{Wasm} has recently emerged as a lightweight alternative for executing external, untrusted code. It provides a robust mechanism for (i) isolating third-party applications, ensuring strong fault containment and fine-grained resource control, and (ii) metering, allowing precise control over application execution and deterministic bounds on resource consumption per application. Furthermore, Wasm ensures portability, dependency management, and resource control akin to containers \cite{wiegratz2024comparing}.
However, adopting Wasm instead of containers involves a trade-off: Wasm provides finer-grained, instruction-level resource control and lower processing overhead in our evaluation, but it may introduce additional latency and memory overhead depending on the runtime configuration and application dependencies. We quantify and discuss these trade-offs in detail in Section \ref{sec:exp}.
\section{dApp Isolation through the Wasm Sandbox} \label{sec:wasm}

\ac{Wasm} emerged as a lightweight sandbox for executing external and untrusted code, offering a modern alternative for hosting third-party applications~\cite{menetrey2022webassembly}. It was originally designed to isolate the execution of applications written in different compiled programming languages (e.g., Rust, C++, and Go) on Web browsers, but it has since been adopted as a general-purpose mechanism to isolate third-party applications, from data centers to embedded devices~\cite{wiegratz2024comparing}. By compiling code from multiple languages into a compact, portable binary format, \ac{Wasm} enables near-native performance while minimizing virtualization overhead.

\subsection{\ac{Wasm} Architecture and Operation}

\begin{figure}[t]
\centering
    \includegraphics[width=1\linewidth]{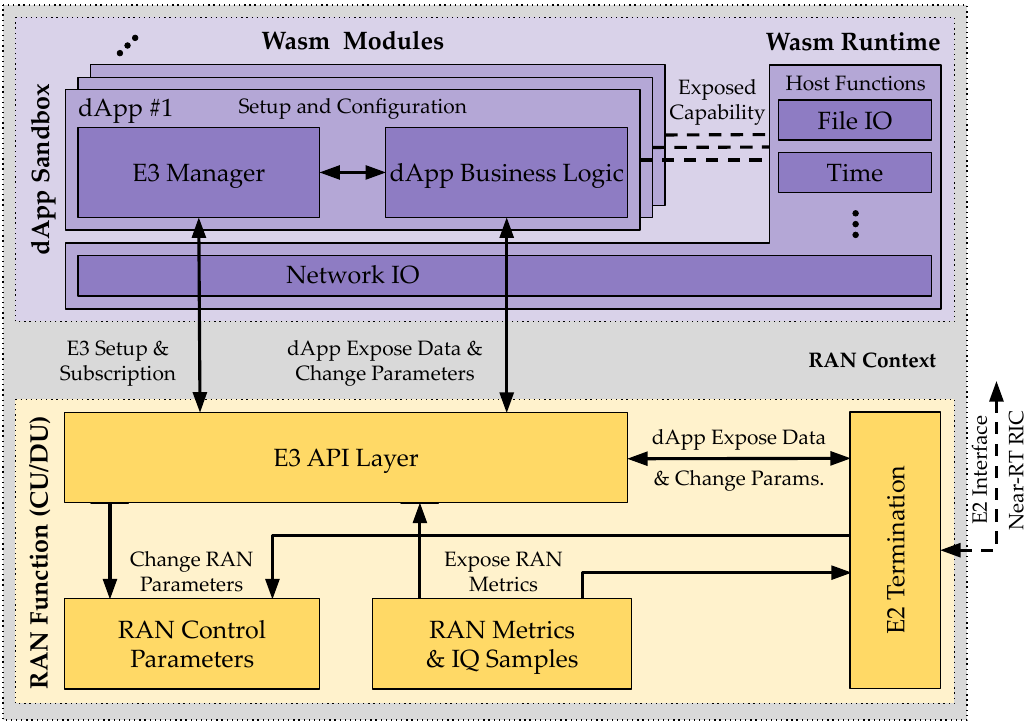}
    \caption{Our architecture to secure the operation of dApps deployed on the \ac{RAN} functions, leveraging \ac{Wasm} to provide isolation and deterministic performance, while running on top of existing frameworks for creating dApps.}
    \label{fig:denegrapp}
\end{figure}

The \ac{Wasm} architecture consists of three key components~\cite{menetrey2022webassembly}: \textit{module}, \textit{runtime}, and \textit{host environment}, as illustrated in the top part of Fig.~\ref{fig:denegrapp}. The \textit{module} encapsulates the business logic of an application, alongside its data and an interface description specifying which functions it imports from or exports to the host. The \textit{runtime} is responsible for loading, validating, and executing these modules within a secure execution environment, providing each with its own memory space and an isolated execution context. Finally, the \textit{host environment} defines the set of \textit{host functions} that modules are allowed to access (akin to system calls), including memory management, I/O operations (e.g., file and network), and timers. It serves as the policy enforcement and resource management layer, controlling resource utilization and providing authentication and authorization for modules.

When a \ac{Wasm} module is deployed, the runtime binds it to the host functions exposed by the host environment~\cite{menetrey2022webassembly}. All module external interactions, e.g., message reception, control command issuance, or measurement storage, must cross these predefined interfaces. These interfaces enforce strict capability-based isolation, such that modules have no direct access to sockets, files, or system calls, and can invoke only operations permitted by the host. To prevent resource exhaustion from individual modules, \ac{Wasm} incorporates execution metering, known as \textit{gas}, to track instruction counts and constrain each module's execution time or computational cost. The gas-based metering is a resource accounting mechanism that associates a predefined computational cost with each Wasm bytecode instruction. Prior to execution, the Wasm module is instrumented to deduct these costs from an allocated gas budget at runtime. When the available budget is exhausted, execution is automatically terminated, providing deterministic bounds on resource usage and preventing infinite loops or other forms of resource abuse. Together, these mechanisms enable multiple dApps to run concurrently while preserving fault containment, security, and timing guarantees, which are key requirements for securing the real-time programmability of \ac{RAN} functions.

\subsection{Opportunities for \ac{Wasm} in \ac{O-RAN}}

\ac{Wasm} offers opportunities for the real-time programmability of RAN functions in O-RAN by providing resource isolation while keeping the control-loop latency within the real-time budget required by dApps. As shown in Section \ref{subsec:controlLoopLatency}, this isolation introduces measurable overhead compared with bare-metal execution; however, the measured control-loop latency remains well below the 10 ms threshold associated with real-time RAN control loops \cite{lacava2025}. Rather than introducing new orchestrators at the edge, we can leverage \ac{Wasm} to provide a common runtime abstraction directly on the \ac{CU}/\ac{DU}, through which multiple \ac{Wasm} modules can be deployed, monitored, and replaced within a \ac{RAN} function. To support \ac{Wasm}, vendors only need to embed a lightweight runtime within their existing \ac{CU}/\ac{DU} software stack and expose a set of host functions tied to controllable parameters and metrics, enabling secure, real-time programmability without modifying the base implementation of the \ac{RAN} function~\cite{wiegratz2024comparing}. In addition, by creating dApps as \ac{Wasm} modules, multiple dApps can coexist safely within a shared infrastructure, each running natively in its own sandbox with resource isolation and predictable performance, while supporting different programming languages.

Adopting \ac{Wasm} unifies the programmability model for developers and standardizes how vendors host external, untrusted code. It provides a standardized interface for creating and deploying dApps across heterogeneous \ac{RAN} implementations. For vendors, this interface enables real-time programmability on their hardware by exposing a single Wasm-compliant layer, simplifying integration and reducing maintenance overhead. For developers, Wasm creates a consistent, platform-agnostic environment for implementing real-time control logic. We can incorporate \ac{Wasm} as a layer on top of the existing frameworks for developing dApps that interact directly with \acp{CU}/\acp{DU}~\cite{lacava2025, d2024dapps}. This combination strengthens security and isolation guarantees while remaining lightweight. \ac{Wasm} extends existing approaches to developing dApps rather than replacing them, providing a practical, incremental path to enable robust real-time programmability of RAN functions in O-RAN with acceptable control-loop latency.

\subsection{dApps as \ac{Wasm} Modules}

The creation of dApps as \ac{Wasm} modules only differs slightly from compiling native binaries, with three main distinctions: \1 the compilation target, \2 the dependency model, and \3 the system interfaces. First, instead of producing machine code for a specific processor architecture, e.g., \texttt{x86\_64} or \texttt{arm64}, the compiler targets the \texttt{wasm32-wasi} architecture, generating a platform-agnostic binary that executes within a \ac{Wasm} runtime. Second, \ac{Wasm} modules must be self-contained, meaning that all dependencies must be compiled into the module. Finally, \ac{Wasm} modules cannot directly invoke system calls, and all interactions with the host environment must be explicitly handled via host functions.

Building on the work of~\cite{d2024dapps} and \cite{lacava2025}, which proposed a dApp architecture that incorporates a novel E3 Agent embedded within the \ac{RAN} functions, this approach extends \ac{O-RAN} with enhanced real-time control and decision-making capabilities. It complements the standard E2 interface by exposing \ac{RAN} metrics and configurable control parameters encoded in ASN.1 over the network, either directly to internal dApps running in the \ac{CU}/\ac{DU}, or to xApps in the \ac{Near-RT RIC} through the E2 Termination, as shown on the bottom part of Fig.~\ref{fig:denegrapp}. Alongside their business logic, each dApp includes an E3 Manager component responsible for setup and registration with the E3 Agent, which acts as a network-facing server that handles messages exchanged between dApps and the underlying \ac{RAN} functions. Detailed information is available in~\cite{lacava2025}.

Since all interactions between dApps and the E3 Agent occur over the network, \ac{Wasm}-based dApps only need permissions to communicate with the E3 Agent and inclusion of dependencies to encode/decode ASN.1 messages. This approach requires only two steps: \1 exposing a minimal set of host functions to handle network operations, such as creating sockets, listening for connections, transferring data, and closing sessions, and \2 embedding an ASN.1 library directly into the dApp module. With very few modifications on the creation of dApps and the exposure of capabilities from the host environment, we can create \ac{Wasm}-based dApps with resource isolation and predictable performance, while maintaining compatibility with existing dApps frameworks~\cite{d2024dapps, lacava2025}. 
\section{Prototyping Wasm-based dApps}\label{sec:poc}

To demonstrate the operation of \ac{Wasm} on top of existing dApp frameworks, we created a reference \ac{Wasm}-based dApp leveraging the dApp framework proposed in~\cite{lacava2025}. In this section, we describe the \ac{RAN} functions used in our prototype, the \ac{Wasm} runtime we adopted for our sandbox environment, the particularities of compiling a dApp into \ac{Wasm}, and the host functions exposed from our host environment.

For our \ac{RAN} functions, we leveraged the \ac{OAI} 5G software stack extended in~\cite{lacava2025, d2024dapps} with the E3 Agent embedded in the \ac{OAI} codebase to support dApps directly on the \ac{CU} and \ac{DU}. The framework for creating dApps proposed in~\cite{lacava2025, d2024dapps} adopts a subscription paradigm and ASN.1 data encoding, akin to the E2 interface in the \ac{O-RAN} specifications~\cite{polese2023understanding}. Through this E3 interface, the dApps register with the E3 Agent embedded into \ac{OAI}, and issue commands following the RIC Services~\cite{santos2025managing} primitive messages (i.e., report, insert, control, policy, and query) for obtaining metrics and controlling parameters from the underlying \ac{RAN} functions, similar to the operation of xApps in the \ac{Near-RT RIC} through the E2 interface, as shown in Fig.~\ref{fig:denegrapp}. Moreover, the E3 Agent supports new primitive messages to enable communication between dApps and xApps, forwarding these messages through the native E2 Agent included in \ac{OAI} software stack towards the \ac{Near-RT RIC}. We have leveraged Wasmtime\footnote{\url{https://wasmtime.dev}}, a lightweight runtime with a small computational footprint, resource isolation guarantees, and portability across different processing architectures. We deployed our \ac{Wasm} runtime collocated with our radio stack, enabling dApps to interact with \ac{RAN} functions over localhost to minimize communication latency.

The authors of~\cite{lacava2025} provide a publicly available dApp for Spectrum Sharing that performs sensing to enable real-time detection of incumbents transmitting in the same spectrum band and notifies the \ac{DU} of \acp{PRB} that must be avoided for scheduling, enabling operation in dynamic spectrum environments. They implemented this dApp in Python, which, unlike compiled languages (e.g., Rust, C++, and Go), requires packaging the interpreter within the Wasm binary, increasing deployment complexity. To avoid this overhead, we reimplemented this dApp to C++, porting its business logic, ASN.1 message formatting, and external interfaces. With the dApp in a programming language with native \ac{Wasm} support, we compiled it as a \ac{Wasm} module using the \textit{wasi-sdk} toolchain with the \textit{wasm32-wasi} target. Our port of the Spectrum Sensing dApp required only two embedded dependencies in the resulting \ac{Wasm} module: a linear algebra library for performing mathematical operations, and an ASN.1 library for encoding/decoding messages. We leveraged \textit{wasi-sdk} for compiling our dependencies from source and bundling them into a \ac{Wasm} portable binary format, our \ac{Wasm}-based dApp.

We configured the host environment to expose basic network host functions for our \ac{Wasm} module to interact with the E3 Agent. Our \ac{Wasm}-based dApp acts as a client, to register and subscribe with the E3 Agent and issue control and query commands, and as a server, listening to incoming report and insert commands. Therefore, we exposed the following host functions to support client and server functionality: \textit{sock\_connect}, \textit{sock\_write}, \textit{sock\_read}, \textit{sock\_bind}, \textit{sock\_accept}, and \textit{sock\_close}. This set of functions serves as a reference point for network interactions, and more complex dApps that require interaction with additional host resources may require additional host functions. Finally, we deployed our \ac{Wasm}-based dApp as a \ac{Wasm} module into the \ac{Wasm} runtime.
\section{Experimental Results}\label{sec:exp}

In this section, we validate the use of \ac{Wasm} as a lightweight sandbox to secure the operation of dApps in \ac{O-RAN}, evaluating its isolation capabilities, impact on real-time control loops, and computational trade-offs.

\subsection{Experimental Setup}

To evaluate the operation of \ac{Wasm}-based dApps, we leveraged the \ac{OAI} 5G software stack using a collocated \ac{CU} and {DU}, extended in~\cite{lacava2025, d2024dapps} with an E3 interface to support hosting dApps. Since our focus is on assessing the isolation and performance impact of different sandbox environments, we configured the radio chain using a direct \ac{RF} loopback, allowing us to examine our dApp under stable and reproducible PHY conditions. To ensure that any performance differences in our experiments are due to overheads from sandboxing environments, and is not limited by the underlying processing platform, we run our experiments on a high-performance computer equipped with an AMD Ryzen 7 5800X processor, 32 GB of RAM, and 1 TB of NVMe storage. To compare the performance of the dApp running in bare metal against different sandboxing environments, we evaluate separate scenarios where we deploy container and \ac{Wasm} runtimes on the same server as our 5G software stack. Specifically, we adopt Docker version 28.3.3 as our container runtime and the WASI SDK 25.0 as our \ac{Wasm} runtime. Further information is available in the repository of our \ac{Wasm}-based dApp (\url{https://github.com/LABORA-INF-UFG/paper-JYPBJACK-2026}).

\subsection{dApp Isolation through \ac{Wasm}}\label{subsec:dAppIsolationThroughWasm}

\begin{figure}[t]
    \centering
    \includegraphics[width=\columnwidth]{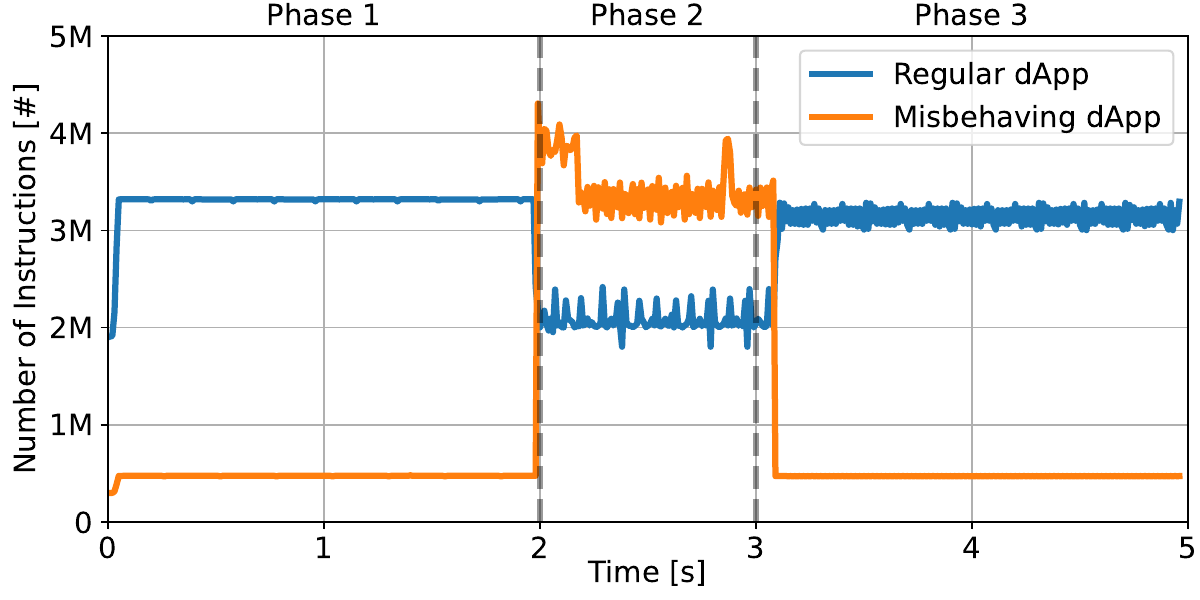}
    \caption{Experiment demonstrating dApps under normal conditions (Phase 1), how misbehaving dApps can compromise the operation of other dApps and their real-time control loops (Phase 2), and the effect of \ac{Wasm}'s fine-grain isolation ensuring predictable performance for dApps (Phase 3). }
    \label{fig:isolation}
\end{figure}

In this analysis, we assess \ac{Wasm}'s ability to isolate dApps from one another, which is critical for running untrusted third-party applications on the mobile network infrastructure, and limiting the impact of misconfigured, faulty, or malicious dApps. To evaluate this scenario, we conducted a controlled experiment with three phases, comprising two artificial \ac{Wasm}-based dApps operating concurrently on a shared \ac{RAN} function: a regular dApp designed to consume approximately 60\% of the CPU instruction rate, and a misbehaving dApp initially designed to consume 20\%. In Phase 1, dApps operate as intended, representing deployments under normal conditions in which multiple dApps coexist without affecting one another. In Phase 2, after \unit[2]{s}, the misbehaving dApp saturates the instruction rate at 100\%, emulating unintentional failures or malicious behavior. In Phase 3, after \unit[3]{s}, we enable \ac{Wasm}'s resource metering and use gas to assign a fixed instruction budget to the regular and misbehaving dApps corresponding to their intended 60\% and 20\% CPU instruction rate, respectively. Figure~\ref{fig:isolation} shows the results of our measurements.

After the initial bootstrap period, we observe a stable CPU instruction rate for dApps in Phase 1. In Phase 2, we observe a spike in the processing demand from the misbehaving dApp, followed by a period where dApps compete for shared resources and fail to meet their intended targets, illustrating how one misbehaving dApp can compromise the operation of other dApps and their real-time control loops. In Phase 3, we observe the effect of Wasm's fine-grained isolation with gas, i.e., each dApp is assigned a fixed instruction budget every \unit[10]{ms} and is suspended once that budget is exhausted until the next window. This mechanism ensures the regular dApp operates close to its intended CPU instruction rate while containing the escalation of the misbehaving dApp. Unlike container-based deployments, which rely on a weighted time slicing of CPU resources to control resource usage~\cite{wiegratz2024comparing}, \ac{Wasm} allows precise control of instructions per application, providing a more predictable and deterministic sandbox for dApps in \ac{O-RAN}.

\subsection{Control Loop Latency}\label{subsec:controlLoopLatency}

\begin{figure}[t]
    \centering
    \includegraphics[width=\columnwidth]{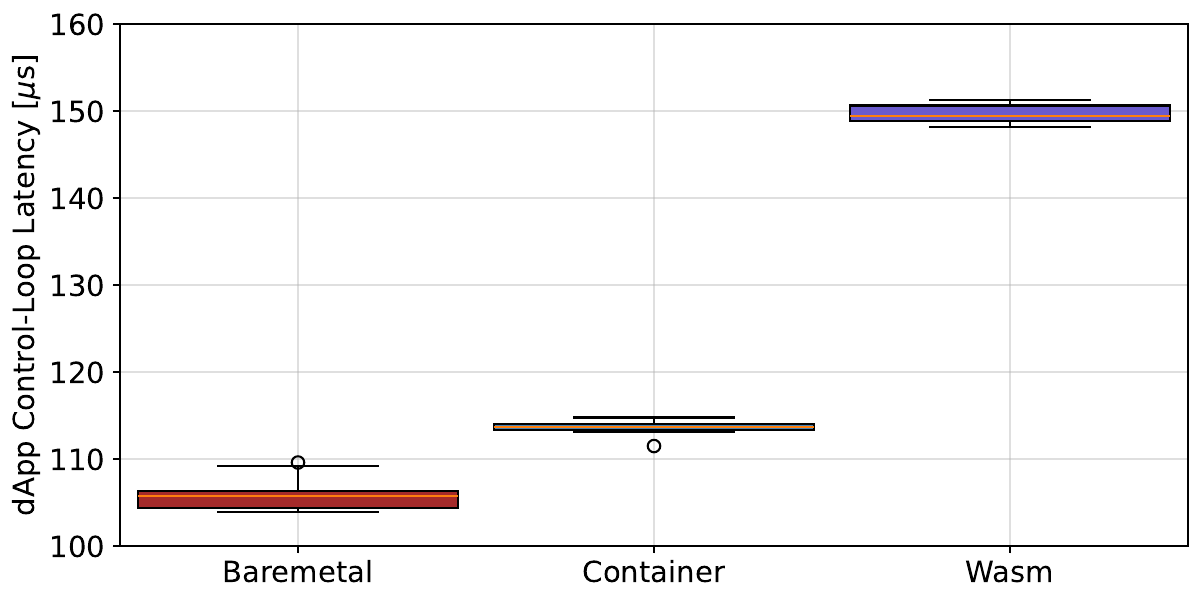}
    \caption{Control loop latency for a dApp running in baremetal, in a container, and in \ac{Wasm} across 10 independent experiments, demonstrating the trade-off between isolation and additional virtualization overhead.}
    \label{fig:times}
\end{figure}

We evaluate the control-loop latency of dApps under different execution environments. As dApps operate within strict real-time constraints to control \ac{RAN} functions, with latencies below \unit[10]{ms}, even small virtualization overheads can significantly impact their performance. Our reference dApp based on~\cite{d2024dapps} performs a Spectrum Sharing control loop that includes decoding ASN.1 messages with data from the \ac{RAN}, analyzing their content to make a decision, encoding the chosen action into ASN.1, and transmitting the information to the E3 agent. We measured the control-loop latency for our reference dApp deployed in bare metal, in a container, and in \ac{Wasm}, as shown in Fig.~\ref{fig:times}. We observe the impact of virtualization overheads when running the dApp in a container and in \ac{Wasm}, with median control-loop latencies of \unit[113.63]{$\mu$s} and \unit[149.44]{$\mu$s}, respectively, corresponding to 7.45\% and 41.31\% higher latency than bare metal, which has a median latency of \unit[105.75]{$\mu$s}. The larger overhead of \ac{Wasm} is due to execution metering, as the \ac{Wasm} runtime tracks executed instructions and periodically compares them against the gas budget assigned to each dApp, representing a trade-off between raw performance and fine-grained isolation provided by \ac{Wasm}. Despite these virtualization overheads, both container- and \ac{Wasm}-based dApps meet the real-time constraints, with control-loop latencies remaining under \unit[10]{ms}, highlighting that we can provide isolation and portability with Wasm-based dApps without violating stringent latency requirements.

\subsection{Computational Footprint}

\begin{figure}[t]
    \centering
    \includegraphics[width=\columnwidth]{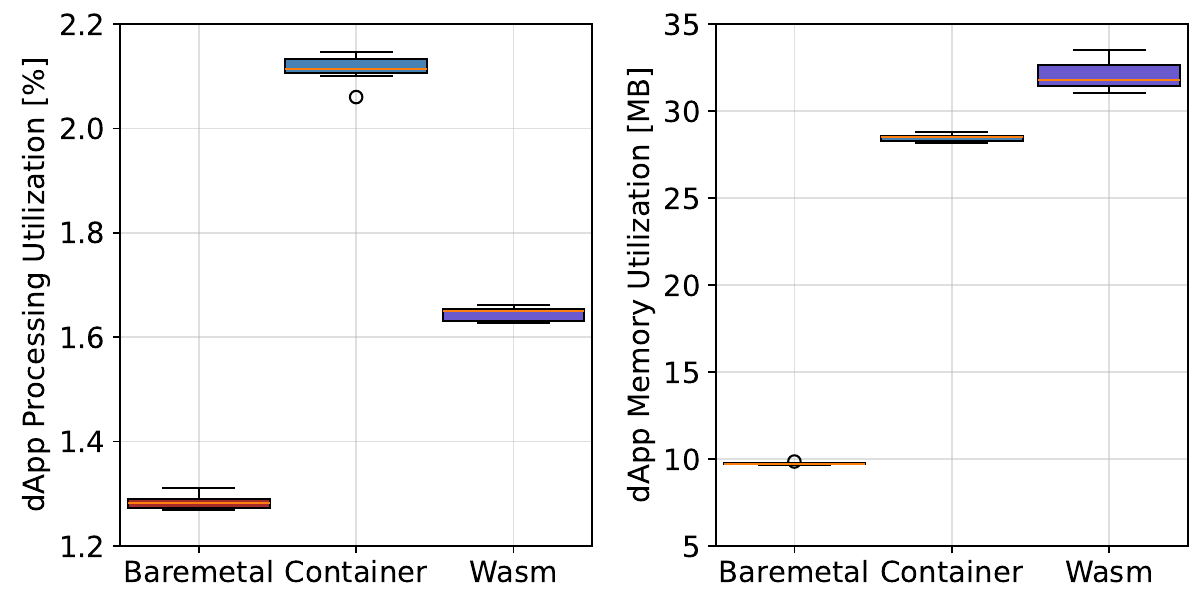}
    \caption{Comparison of the computational footprint for a dApp running in bare metal, in a container, and in Wasm in terms of processing and memory utilization across 10 independent experiments. Processing utilization corresponds to the CPU usage reported by \texttt{psrecord}\textsuperscript{\thepsrecordfn} for the monitored dApp process tree, sampled at 1~s intervals under fixed CPU-affinity conditions.}
    \label{fig:resources}
\end{figure}

We analyze the computational footprint of running dApps under different execution environments. This evaluation allows us to quantify the impact of different sandboxing environments on resource consumption, which is important in edge deployments, where computing resources are typically scarce and costly. We measured the processing and memory utilization for our reference dApp deployed in bare metal, in a container, and in \ac{Wasm}, as shown in Fig.~\ref{fig:resources}. The processing utilization corresponds to the CPU usage reported by \texttt{psrecord}\footnote{\label{fn:psrecord}https://github.com/astrofrog/psrecord}\newcounter{psrecordfn}\setcounter{psrecordfn}{\value{footnote}} for the monitored dApp process and its child processes, sampled at 1~s intervals. In our experiments, the monitored process was pinned to a fixed CPU set using \texttt{taskset}, ensuring that all deployment scenarios were measured under the same CPU-affinity conditions.

We observe the impact of virtualization overheads in containers and \ac{Wasm} relative to bare metal. In terms of processing, containers introduce a median of 64.84\% processing overhead, while \ac{Wasm} reduces this overhead to 28.91\%. In terms of memory, containers and \ac{Wasm} consume $2.93\times$ and $3.28\times$ than bare metal, respectively, which has a median memory footprint of 9.70 MB. These results demonstrate that neither sandboxing environment is uniformly lighter across all resources. Compared with containers, Wasm reduces processing utilization by 21.80\%, but incurs 11.51\% additional memory overhead. Therefore, the main advantage of \ac{Wasm} is combining lower processing utilization with deterministic per-application resource control through gas-based instruction metering. As shown in Section \ref{subsec:dAppIsolationThroughWasm}, this mechanism assigns each dApp a fixed instruction budget and suspends its execution once the budget is exhausted, whereas container-based deployments rely on weighted CPU time-slicing and do not provide the same instruction-level control.

\section{Conclusions and Open Challenges}\label{sec:conc}

In this paper, we explored opportunities for \ac{Wasm} in \ac{O-RAN} as a lightweight, secure sandbox environment for dApps, and demonstrated a proof-of-concept \ac{Wasm}-based dApp. \ac{Wasm} provides capabilities beyond container-based deployments, enabling fine-grained isolation and deterministic performance through instruction metering, while being less computationally expensive in terms of processing. However, there are many open challenges for research and standardization communities, beyond those addressed in this work. 

As with any sandbox environment, \ac{Wasm} introduces a new layer of abstractions and complexity in the mobile network, which incur overheads for dApps, manifested as latency in their control loops. This latency may affect the execution of their real-time business logic and degrade the performance of \ac{RAN} functions. This challenge creates an opportunity for future works to evaluate the impact of control loop latency of \ac{Wasm}-based dApps on \ac{RAN} performance metrics, and the potential benefits of optimization techniques such as \ac{JIT} and \ac{AOT} compilation to improve dApp performance. 

A deeper analysis of latency overhead still requires further investigation, such as identifying its sources (e.g., metering, runtime boundary crossings) and potential optimization strategies. Carrier-grade mobile networks deploy hardware accelerators at the edge, such as FPGAs, GPUs, and SmartNICs, to meet the performance requirements of the 5G radio stack. While \ac{Wasm} offers portability across different processor architectures, its support for computing platforms is still in its infancy, with ongoing standardization efforts and few prototypes exploring hardware acceleration~\cite{ramesh2025empowering}. In this context, \ac{Wasm}-based dApps may not be able to operate at the same scale as 
their native counterparts. Nevertheless, hardware acceleration for third-party applications in \ac{O-RAN} is a largely underexplored topic.

Practical deployment also raises operational challenges beyond the runtime overheads measured in this work. \ac{Wasm} supports flexible software packaging, allowing applications to be deployed either as standalone images or as modular compositions. The choice of packaging strategy involves a trade-off among deployment simplicity, incremental updates, and runtime initialization overhead. Operators also need mechanisms for safely updating \ac{Wasm} modules, validating new versions before activation, and rolling back faulty deployments. In addition, the host functions and E3 APIs exposed by the CU/DU must be versioned, so that dApps compiled against one runtime interface remain compatible with future runtime, RAN-function, and API updates.

Finally, a comprehensive security analysis is needed to formalize the threat model (e.g., malicious or faulty dApps) and compare the security of Wasm-based isolation with container-based approaches. From a system perspective, it is also critical to understand how \ac{Wasm}-based dApps will interact with external orchestration mechanisms in the \ac{Near-RT RIC} and the \ac{SMO}, to inform ongoing standardization efforts and assess alignment with future \ac{O-RAN} specifications, including the potential inclusion of \ac{Wasm} sandbox environments in the O-Cloud. In future works, we plan to investigate the storage impacts of compiling dependencies of Wasm-based dApps, and the scalability of large-scale, multi-dApp scenarios.

\bibliographystyle{IEEEtran}
\bibliography{references.bib}

\section*{Acknowledgment} 
This work was supported by the CNPq under grant number 306283/2025-5, by the MCTIC/CGI.br/FAPESP under grants Nos. 2020/05127-2, 2020/05182-3 and 2025/01970-0, by the OpenRAN Brazil project under grant A01245.014203/2021-14, and by CAPES. This work also received support from the National Science Foundation US-Ireland R\&D Partnership program under grant No. 2421362. The research leading to this article was also supported by the Commonwealth Cyber Initiative.

\newpage
\begin{IEEEbiographynophoto}{Jo\~ao Paulo Esper} (joaopauloesper@discente.ufg.br)
is a PhD student in Computer Science at UFG.
\end{IEEEbiographynophoto}

\begin{IEEEbiographynophoto}{Yure Freitas}(yuregabriel@discente.ufg.br)
is currently pursuing a B.Sc. in Computer Science at UFG.
\end{IEEEbiographynophoto}

\begin{IEEEbiographynophoto}{Pedro Souza}(pedromasou@edu.unisinos.br)
is a researcher at UNISINOS.
\end{IEEEbiographynophoto}

\begin{IEEEbiographynophoto}{Bruno Silvestre}(brunoos@ufg.br)
is an Associate Professor at UFG.
\end{IEEEbiographynophoto}

\begin{IEEEbiographynophoto}{Joao F. Santos}(joaosantos@vt.edu, Senior Member, IEEE)
is a research assistant professor with CCI at VT. 
\end{IEEEbiographynophoto}

\begin{IEEEbiographynophoto}{Alexandre Huff}(alexandrehuff@utfpr.edu.br)
is an Associate Professor at UTFPR.
\end{IEEEbiographynophoto}

\begin{IEEEbiographynophoto}{Cristiano Both}(cbboth@unisinos.br)
is a professor at UNISINOS.
\end{IEEEbiographynophoto}

\begin{IEEEbiographynophoto}{Kleber Cardoso} (kleber@ufg.br) 
is a Full Professor at UFG.
\end{IEEEbiographynophoto}

\end{document}